\documentclass[12pt,a4paper]{article}

\usepackage{latexsym}
\usepackage{amsmath}
\usepackage{amsfonts}
\usepackage{amssymb}
\usepackage{amsthm}
\usepackage{amscd}
\usepackage{mathrsfs}

\newcommand{\be}{\begin{equation}}
\newcommand{\ee}{\end{equation}}
\newcommand{\bea}{\begin{eqnarray}}
\newcommand{\eea}{\end{eqnarray}}

\def\ri{{\mathrm{i}}}                   %
\def\cL{{\cal L}}                       %
\def\bR{{\mathbb R}}                    %
\def\1{{\mbox{\boldmath $1$}}}          %
\def\tr{\mathrm{tr\,}}                  %
\def\0{{\mbox{\boldmath $0$}}}          %
\def\jp{\frac{1}{2}}
         
\def\si{\sigma}               
\def\cL{{\cal L}}                        %

\def\e{\epsilon}                    %

\def\no{\noindent}

\def\om{\omega}
\def\w{\wedge}
\begin{document}

\begin{center}
{\Large \bf  
Totally classical Calogero model} \end{center}

\vspace{0.2cm}

\begin{center}
 C. Klim\v c\'\i k   \\

\bigskip

 Institut de math\'ematiques de Luminy,
 \\ 163, Avenue de Luminy, \\ 13288 Marseille, France\\
 e-mail: ctirad.klimcik@univ-amu.fr

\bigskip

\end{center}

\vspace{0.2cm}

\begin{abstract} We    show that  the standard Calogero Lax matrix can be interpreted as a function on the fuzzy sphere
and the Avan-Talon $r$-matrix as a function on the direct product of two fuzzy spheres.  We calculate the  limiting Lax function and $r$-function  when the fuzzy sphere tends to the ordinary sphere and we show that they  define an integrable model
interpreted as   a large $N$ Calogero model  by Bordemann, Hoppe and Theisen.

 \end{abstract}

\newpage

\section{Introduction}

The formalism based on the concepts of the Lax matrix and the Lax pair  turns out to be very efficient in a study of a large class of classical
completely integrable models
\cite{A,BBT,BV,Lax}.  Briefly speaking, the Lax pair is a pair of matrix valued functions $L,M$ on the phase space $P$ of the integrable
model such that the equations of motion of the latter can be written in terms of the matrix commutator
  \be \frac{d}{dt}L=[L,M]. \label{mla}\ee
 Moreover,  the Lax matrix $L$ is required to generate integrals of motion in Poisson involution, i.e. it should fulfil a condition
 \be \{\tr L^m,\tr L^n\}_P=0, \quad \forall m,n.\label{fin}\ee
Bordemann, Hoppe and Theisen  \cite{BHT} had an idea to generalize this formalism by replacing the 
  matrices $L,M$    by functions on some two-dimensional auxiliary compact symplectic manifold $B$ and the commutator by  the auxiliary Poisson bracket $\{.,.\}_B$:
\be \frac{d}{dt}L=\{L,M\}_B.\label{plp}\ee
   Moreover,  they  imposed the following condition on $L$
 \be\biggl\{\int_B\omega L^m,\int_B\omega L^n\biggr\}_P=0, \quad \forall m,n,\label{ia}\ee 
 where $\omega$ denotes the auxiliary symplectic form on $B$.
 
 \no Starting with a suitable ansatz for the Lax function, Bordemann, Hoppe and Theisen \cite{BHT}  have identified  several field theories in $1+1$ dimensions which
 are integrable via their novel mechanism. Among them they singled out two models  the classical actions  of which read respectively  \be S_t=\jp\int dt\int_{-1}^{1} d\sigma
\biggl(\dot{q}(\si)^2-a^2\exp{{q'(\si)}}\biggr),\label{1}\ee
 \be S_c =\jp\int dt\int_{-1}^{1} d\sigma
\biggl(\dot{q}(\si)^2-\frac{a^2}{{q'(\si)}^2}\biggr).\label{2}\ee
Here  the dependence of the dynamical field $q(\si)$ on the time $t$ is tacitly understood,   $a$ is a coupling constant, 
the "dot" and "prime" denote the time and the space derivatives, respectively, and   boundary conditions on $q(\si)$ will be specified in Section 3. 

\no It is instructive to compare the field theories (\ref{1}),(\ref{2}) with the integrable $N$-particle  Toda and Calogero systems:
\be S_{tN}=\jp\int dt \biggl(\sum_{1\leq i\leq N}
\dot{q}_i^2-\kappa_N^2\sum_{1\leq  i \leq N}\exp{(q_i-q_{i+1})}\biggr),\label{1N}\ee
\be S_{cN}=\jp\int dt \biggl(\sum_{1\leq i\leq N}
\dot{q}_i^2-\sum_{1\leq i\neq j\leq N}\frac{\kappa_N^2}{(q_i-q_j)^2}\biggr),\label{2N}\ee
 The similarity between field theoretical and $N$-particle expressions was  not the only reason why Bordemann, Hoppe and Theisen   interpreted the  models   (\ref{1}),(\ref{2}) as   "infinite dimensional analogues of $N$-particle Toda- and Calogero  systems". In fact, 
 although they did not derive  these field  theories as  large $N$ limits
 of  the $N$-particle models, they  argued that it should be possible to do it 
 because   the auxiliary Poisson algebra of functions on the two dimensional compact symplectic manifold $B$ can be viewed as infinite-dimensional version   $gl(\infty)$ of the algebra  $gl(N)$ of finite size $N\times N$-matrices  \cite{Ho1}.   The parameter $N/2$ should be then interpreted   as the Planck constant of an auxiliary quantization converting the auxiliary Poisson algebra into the matrix algebra 
 and the integrable field theories (\ref{1}),(\ref{2}) could be therefore called "doubly classical" or "totally classical".
 
 \no The suggestion of \cite{BHT} to obtain the Lax functions of the totally classical field models (\ref{1}),(\ref{2})  as a large $N$ limit
 of 
 the known  
 Lax matrices of the $N$-particle models (\ref{1N}),(\ref{2N}) was then realized  by  Hoppe in the Toda case \cite{Ho3}.   However, to   our best knowledge,  a similar result was not obtained
 for the Calogero model. 
 We wish to stress that the similarity between the field theoretical model and the finite $N$-particle theory is more evident  for the Toda model
 than for the Calogero one. Indeed,   the  Toda field theory (\ref{1}) is  local in the sense that its Lagrangian contains only
 the first derivative $q'(\si)$ and the   discrete theory (\ref{1N})  is also local in the sense that the potential contains only the
 interaction of the nearest neighbours. In reality,  the $N$-particle Toda model (\ref{1N}) is nothing but  a straightforward discretization of the Toda field theory
 (\ref{1}).

\no  In the Calogero case the situation is much more subtle,  because the  field theory (\ref{2}) is local but the $N$-particle Calogero model
 (\ref{2N}) is non-local. Indeed, the Calogero model (\ref{2N}) is not a discretization of the field theory (\ref{2}). On the top of it, there
exists 
a  {\it non-local}  field theory   referred to
 as the large $N$ Calogero model in the literature \cite{ABW,PO} the discretization of which does give the $N$-particle Calogero model (\ref{2N}). Would 
  the existence of two different large $N$ limits of the same finite $N$ theory   indicate  a contradiction?  A priori no, 
since much depends in which context the large $N$ limit is taken, what other parameters get fixed while taking the limit etc. \cite{ A2,CCK, FLSZ,GN1,JS,KY,MP,O1}. 
We shall  actually
justify the interpretation of  the model (\ref{2})  as one   particular large $N$ limit  
  by appropriately tuning the coupling constant $\kappa_N$  and the Darboux symplectic $\Omega_N$  form of the 
$N$-particle Calogero model (\ref{2N}).   Speaking more precisely, we shall prove that the discrete  non-locality gets eliminated
and the finite $N$ Calogero model approaches  the totally classical field theory (\ref{2}) if we make 
 both  $\kappa_N$ and $\Omega_N$ proportional  to  $1/N$.  This will constitute the first   result of  the present paper.
  
  \no In the original article \cite{BHT}, the fundamental involutivity relation (\ref{ia}) was obtained by a direct computation starting 
  from an appropriate ansatz for the Lax function.  In the finite $N$ case, the involutivity (\ref{fin}) can be proved most easily by
  using the concept of the classical $r$-matrix. We recall  that the classical $r$-matrix  \cite{BBT,BV} is a map $r(N)_{12}$  from the phase space of the
  model into the direct product of two copies of the matrix algebra such that a fundamental compatibility relation with the Lax matrix
  $L(N)$ holds:
\be \{L(N)_1,L(N)_2\}_{P}=-\frac{\ri N}{2} [r(N)_{12},L_1(N)]+\frac{\ri N}{2} [r(N)_{21},L_2(N)],\label{stl}\ee
where $L(N)_1=L(N)\otimes 1$ and $L(N)_2=1\otimes L(N)$. It is then very easy to show that (\ref{stl}) implies (\ref{fin}).

\no In the totally classical context, Hoppe \cite{Ho2} has suggested to introduce a concept of an $r$-function on $P\times B\times B$, which would
verify an analogue of (\ref{stl}):
 \be \{L(z),L(w)\}_P=\{r(z,w),L(z)\}_B-\{r(w,z),L(w)\}_B.\label{rL}\ee
Here $L$ is the Lax function,  the dependence of $L$ and $r$  on the coordinates of $P$ is tacitly understood in (\ref{rL}) 
and the letter $z$ or $w$ stands for some parametrization of the auxiliary symplectic manifold $B$. As in the finite $N$ case it is easy
to show
that the involutivity relation (\ref{ia}) is the consequence  of (\ref{rL}). 

\no In the case of  the totally classical Calogero model, Hoppe started to look for  the $r$-function  by  choosing an appropriate ansatz which he then substituted into the condition (\ref{rL}).  
Remarkably,  the resulting object  which he has found is not quite an $r$-function but rather an $r$-distribution  in the variables parametrizing
  $B\times B$!  This may look surprising at a first sight: how the large $N$-limit  may convert matrices into distributions?   As we shall see, this is indeed the case and the $r$-distribution does arise in the large $N$ limit. The  
  derivation  of  Hoppe's $r$-distribution directly from the Avan-Talon $r$-matrix \cite{AT} of the $N$-particle Calogero model
  constitutes the second  result of the present article.

\no In this paper we choose the ordinary sphere as the auxiliary symplectic manifold $B$ and    in Section 2 we review its quantization called the fuzzy sphere \cite{Ho1,Mad}.   Then in Sections 3 and 4 we prove that the Lax matrix and $r$-matrix of the $N$-particle Calogero
model are respectively the fuzzy quantizations of the Lax function   \cite{BHT} and of the $r$-distribution \cite{Ho2}  of the totally classical
Calogero model (\ref{2}).
  We finish by a short outlook.

\section{The fuzzy sphere}
The ordinary symplectic sphere $S^2$ is a surface embedded in $\bR^3$ defined in Carthesian coordinates $x_1,x_2,x_3$ as
\be x_1^2+x_2^2+x_3^2=1.\ee
 We parametrize it as 
 \be x_1=\sqrt{1-\si^2}\cos{\phi},\quad x_2=\sqrt{1-\si^2}\sin{\phi},\quad x_3=\si, \qquad  \sigma\in [-1,1],\phi\in [-\pi,\pi].\label{parr}\ee
 The standard round symplectic form on $S^2$ is then given by
\be \omega\equiv -\frac{1}{2}\e_{jkl}x_j dx_k\wedge dx_l\bigg\vert _{S^2}= d\si\w d\phi.\label{spsy}\ee
 The result of the quantization of the symplectic manifold $(S^2,\omega)$ is known   as 
 "the fuzzy sphere" \cite{Ho1,Mad}.  The   linear $SO(3)$-equivariant quantization map $Q_N$   associates  to  smooth functions $f$ on $S^2$   sequences  
of $N\times N$-matrices  $Q_N(f)$ which are called the quantized or fuzzy functions.    We shall not need an explicit  formula for the  quantization map $Q_N$ but we do need three basic properties of it:
\be Q_N(f)Q_N(g)=Q_N\biggl(fg+O\biggl(\frac{2}{N}\biggr)\biggr),\label{p1}\ee
\be [Q_N(f),Q_N(g)]=Q_N\biggl(\ri\frac{2}{N}\{f,g\}_B+ O\biggl(\frac{4}{N^2}\biggr)\biggr),\label{p2}\ee
 \be  \frac{1}{2\pi}\int_{S^2}\omega f = \tr Q_N\biggl(\frac{2}{N}f+O\biggl(\frac{4}{N^2}\biggr)\biggr).\label{p3}\ee
 Obviously the parameter $2/N$ plays the role of the auxiliary  Planck constant.
 
\no To give a flavor, what the map $Q_N$ is about, let us  make explicit  the quantized versions of the   functions  $x_3$,
$x_1\pm \ri x_2$   defined in (\ref{parr}):
\be Q_N(x_3)_{ij}=\frac{1}{\sqrt{N^2-1}}(N+1-2j)\delta_{ij},\quad Q_N(x_1+ \ri x_2)_{ij}=\frac{2}{\sqrt{N^2-1}} \sqrt{(j-1)(N-j+1)}\delta_{i,j- 1} \label{fux}\ee 
and  $Q_N(x_1- \ri x_2)$ is the Hermitian-conjugated matrix  $Q_N(x_1+ \ri x_2)^\dagger$. In particular, it is then easy to verify that it holds the emblematic fuzzy sphere relation 
\be Q_N(x_1)^2+ Q_N(x_2)^2+ Q_N(x_3)^2=\1_N,\label{fuz}\ee
where $\1_N$ stands for the unit $N\times N$-matrix.  

\no So far we have learned  that every smooth function on the sphere gives  rise to a sequence of $N\times N$-matrices.
 It is perhaps less known that appropriate sequences of $N\times N$-matrices may represent  quantizations of not  just  smooth functions, but also  of singular functions and/or  distributions
on $S^2$. As an example particularly relevant for the present paper, we now describe the fuzzy   vortices. For that, consider
the sequence of $N\times N$ matrices $V(N)$ defined by their matrix elements:
\be V(N)_{ij}=\delta_{i,j- 1},\quad i,j=1,...,N.\ee
In words: $V(N)$ is the Jordan block with zeros on the principal diagonal.
We immediately check that
\be Q_N(x_1+\ri x_2)= \sqrt{1-\frac{\sqrt{N-1}}{\sqrt{N+1}}Q_N(x_3)}\sqrt{1+\frac{\sqrt{N+1}}{\sqrt{N-1}}Q_N(x_3)}V(N)\ee
and we note that  the diagonal matrices of which we take the square roots have all eigenvalues  strictly positive. 
In the limit $N\to\infty$, we then obtain
\be x_1+\ri x_2= \sqrt{1-x_3^2}\  V_\infty,\ee
which together with (\ref{parr}) leads to  an identification of $V_\infty$ with the ordinary sphere vortex configuration $e^{\ri\phi}$ and to
an extension of the quantization map to this vortex configuration by setting
\be Q_N(e^{\ri\phi})=V(N).\label{vor}\ee
Another relevant   quantized  singular object on the   sphere is characterized by the following sequence of $N\times N$ matrices $K(N)$:
\be K(N):= \sum_{k,l}^N E_{kl},\label{KN}\ee
where $E_{kl}$ is the elementary matrix with $1$ in $k^{th}$ row and  $l^{th}$ column and $0$ everywhere else.
Note that $K(N)$ are  matrices with all elements   equal to $1$.  
For a finite $N$, we observe
\be K(N)= (V(N)^\dagger)^{N-1}+...+(V(N)^\dagger)^2+V(N)^\dagger+\1_N+V(N)+V(N)^2+...+V(N)^{N-1}  \ee
 which gives from (\ref{vor})
 \be K_\infty={\rm lim}_{N\to\infty} Q_N\biggl(\sum_{j=1-N}^{N-1}e^{\ri j\phi}\biggr)=2\pi\delta(\phi).\ee
 Said differently, we extend the quantization map to the delta function by setting
\be 2\pi Q_N(\delta(\phi))=K(N).\label{del}\ee
Now we  use the same symbol $Q_N$ for the quantization of the direct  product $S^2\times S^2$.  We can argue that
\be Q_N\biggl(\delta(\si_1-\si_2)\delta(\phi_1-\phi_2)\biggr)=\frac{N}{4\pi}\sum_{k,l}^N E_{kl}\otimes E_{lk}.\label{tif}\ee
Indeed, the $\delta$-function $\delta(\si_1-\si_2)\delta(\phi_1-\phi_2)$ on $S^2\times S^2$ is characterized by the property
\be   \int_{-1}^1 d\sigma_2 \int_{-\pi}^\pi d\phi_2\delta(\si_1-\si_2)\delta(\phi_1-\phi_2) f(\si_2,\phi_2)=\int \om_2\delta(\si_1-\si_2)\delta(\phi_1-\phi_2) f(\si_2,\phi_2)=  f(\sigma_1,\phi_1),\label{dell}\ee
where $f(\si_2,\phi_2)$ is an arbitrary smooth function on $S^2$. Following (\ref{p3}), the quantized version of (\ref{dell}) is
\be \frac{4\pi}{N}\tr_2\biggl(Q_N(\delta(\si_1-\si_2)\delta(\phi_1-\phi_2))(\1_N\otimes Q_N(f))\biggr)= Q_N(f).\ee
On the other hand,  for  the  $N\times N$-matrix $Q_N(f)$ it obviously holds
\be  \tr_2\biggl((\sum_{k,l}^N E_{kl}\otimes E_{lk})(\1_N\otimes Q_N(f))\biggr)=Q_N(f).\ee
Comparing the last two equalities   leads to the identification (\ref{tif}).

\no Finally, we shall need also the fuzzy version of the "diagonal" delta function $\delta(\si_1-\si_2)$ on $S^2\times S^2$.  Since the fuzzy version of a $\phi$-independent function on $S^2$ is a diagonal matrix,  the fuzzy version of $\delta(\si_1-\si_2)$ must be given as a sum of   direct products of diagonal matrices. To find out which diagonal matrices appear in those  direct products
we look for a fuzzy analogue of the following identity:
 \be \frac{1}{2\pi}\int_{-1}^1 d\sigma_2 \int_{-\pi}^\pi d\phi_2\delta(\si_1-\si_2)  f(\si_2)=\frac{1}{2\pi}\int\om_2\delta(\si_1-\si_2)  f(\si_2)=  f(\sigma_1).\label{del2}\ee
 It obviously reads 
 \be \frac{2}{N}\tr_2\biggl(Q_N(\delta(\si_1-\si_2))(\1_N\otimes Q_N(f))\biggr)=Q_N(f).\ee
We have also the following  matrix identity:
 \be   \tr_2\biggl((\sum_{k}^N E_{kk} \otimes E_{kk })(\1_N\otimes Q_N(f))\biggr)=Q_N(f),\label{fd}\ee
 which holds for the diagonal matrix $Q_N(f)$. 
 Comparing  the last two equalities leads to the desired quantization formula
 \be  Q_N(\delta(\si_1-\si_2))=\frac{N}{2}\sum_k^N E_{kk}\otimes E_{kk}.\label{ddf}\ee

\section{Large $N$ limit of the Calogero Lax matrix} 
\no Recall first  the standard formula for the Lax matrix $L(N)$ of the Calogero model \cite{OP1}
\be L(N)_{ij} =p_i\delta_{ij} +(1-\delta_{ij})\frac{i\kappa_N}{q_i-q_j}, \quad 1\leq i,j,\leq N,\label{NLax}\ee
where $p_j,q_j$ are the Darboux coordinates on the phase space of the model. In what follows, we  normalize the Darboux symplectic structure $\Omega_N$ and   the coupling constant $\kappa_N$  as 
\be \Omega_N=\frac{2}{N} dq_j\w dp_j,\quad \kappa_N =\frac{ c}{N}.\label{scn}\ee
We now  introduce    important diagonal $N\times N$-matrices $R(N)$  and $P(N)$ as follows
\be R(N)_{ij}:=q_i\delta_{ij}, \quad  P(N)_{ij}:=p_i\delta_{ij}, \quad i,j=1,...,N \label{777}\ee
and we naturally interpret $R(N)$ as $Q_N(q(\si))$ and $P(N)$ as $Q_N(p(\si))$.
 Speaking more precisely, we have  from (\ref{fux})  
\be q_j\equiv  Q_N(q(\si))_{jj} =q\biggl(\frac{N+1-2j}{\sqrt{N^2-1}}\biggr)\label{seq}\ee
and similarly for $p_j$.

\no Few comments are in order to justify the interpretation of $R(N)$ as $Q_N(q(\si))$ and $P(N)$ as $Q_N(p(\si))$. First of all,
$q(\si),p(\si)$ are viewed as the functional coordinates of 
the  {\it field theoretical} phase space   of the totally classical Calogero model (\ref{2}) with the Darboux Poisson bracket
\be \{p(\sigma_1),q(\sigma_2)\}_P=\delta(\sigma_1-\sigma_2), \label{fdb}\ee but at the same time they are viewed as
$\phi$-independent functions on the sphere.   
We note also that for $j$ running through the set  $1,...,N$ the argument $\frac{N+1-2j}{\sqrt{N^2-1}}$ in (\ref{seq}) runs equidistantly  through the interval
$[-1,1]$ which is indeed the domain of definition of the function $q(\si)$. Moreover, for neighbouring $j$ and $j+1$ the distance
of the arguments is $\frac{2}{\sqrt{N^2-1}}$  which for large $N$ is nothing but the $B$-Planck constant.  We thus observe
that the phase space $P$ of the totally classical Calogero model (\ref{2}) gets drastically shrunk by  the  fuzzification. Indeed, among
all functions $q(\si),p(\si)$ which constitute points in $P$, only   functions  constant on  the equidistant intervals
of  the lenghts $\frac{2}{\sqrt{N^2-1}}$ survive the  fuzzification and form a $2N$-dimensional phase space $P(N)$ of the
$N$-particle Calogero  model. Moreover, we can derive  from the Darboux Poisson bracket (\ref{fdb}) on  the phase space $P$ 
 the Poisson bracket on the $2N$-dimensional phase space $P(N)$ parametrized by $q_j,p_j$, $j=1,...,N$. To do that, we consider a function
$T(\si)$ on the sphere and its diagonal quantization $Q_N(T)$. From  the Darboux Poisson structure  (\ref{fdb}), we obtain
\be \biggl\{p(\si),\frac{1}{2\pi}\int \om Tq\biggr\}_P=  T(\si).\label{dkk}\ee
Now from the property (\ref{p3}) we see that the fuzzification  of the formula (\ref{dkk}) must   give
\be \biggl\{p_i,\frac{2}{N}\tr (Q_N(T)R(N))\biggr\}_{P(N)}=  Q_N(T)_{ii}.\ee
We thus infer 
\be \{p_i,q_j\}_{P(N)}=\frac{N}{2}\delta_{ij}\label{jef}\ee
which is consistent with the first  formula  in (\ref{scn}).

 \no Let us verify that the $N$-particle Calogero Lax matrix $L(N)$ (\ref{NLax}) has as a  large $N$-limit   some classical Lax observable $L$.
Said differently, we shall show that $L(N)$
can be interpreted  as  the fuzzification $Q_N(L)$ of some function $L$ on the ordinary sphere.

\noindent It is now easy to verify  with the help of  (\ref{del})  that
 \be [R(N) , L(N)]=\ri\kappa_N(K(N)-\1_N)=\frac{\ri c}{N}Q_N(2\pi\delta(\phi)-1),\label{cos}\ee
where the matrix $K(N)$ was defined in (\ref{KN}).  If it is true that $L(N)$  can be identified with  $Q_N(L)$ for some $L$ then we could rewrite (\ref{cos}) as 
  \be [Q_N(q(\si)), Q_N(L)]= \frac{\ri c}{N}Q_N(2\pi\delta(\phi)-1).\label{cos2}\ee
 which, due to  the fundamental quantization property (\ref{p2}),  would lead  to
 \be \{q(\si),L(\si,\phi)\}_B= -q'(\sigma)\partial_\phi L= \frac{c}{2}(2\pi\delta(\phi)-1).\label{alm}\ee
And indeed!  The differential condition (\ref{alm}) has the following obvious solution
\be L(\si,\phi)=p(\sigma)+\frac{c} { 2\rho'(\sigma)}( \phi-\pi{\rm sign}(\phi)).\label{lax1}\ee
which coincides\footnote{The comparison of (\ref{lax1}) with the result of \cite{BHT} must take into account the range of the parameter 
$\phi$ which is $[-\pi,\pi]$ in our paper and it was $[0,2\pi]$ in \cite{BHT}. We have opted for the different range in order to stress that the Lax function $L$ is discontinuous. This  fact is indeed less visible in \cite{BHT} since the discontinuity occurs precisely on the boundaries
of the range $[0,2\pi]$.}  for $c=2\sqrt{3}a/\pi$   with the Lax function of the   model (\ref{2}) as found in \cite{BHT}.  Moreover, 
 the  condition (\ref{alm})  determines the function $L$ almost unambiguously.  The only ambiguity consists
in adding   a $\phi$-independent function  to the solution   (\ref{lax1}), however, this ambiguity is fixed by the fact 
 that the diagonal term
$p_i\delta_{ij}$ in  the Lax matrix $L(N)$    must be  equal to  $Q_N(p(\sigma))$. 
Thus  we have  justified  the interpretation \cite{BHT}  of the totally classical Calogero model 
as the large $N$-limit  of the $N$-particle Calogero model.

 \no {\bf Remark}:
 It is important to stress that, as it stands, the formula (\ref{lax1}) defines only a function on a subset of $S^2$ covered by the
coordinate chart $(\sigma,\phi)$ and we need   the classical Lax observable everywhere on $S^2$. Fortunately, the function
$\phi-\pi{\rm sign}(\phi)$ smoothly extends  to the anti-Greenwich meridian $\phi=\pm \pi$ and we can also
extend $L_\infty(p,\rho;\sigma,\phi)$ to the poles $\sigma=\pm 1$  if we impose the following boundary conditions \be {\rm lim}_{\sigma\to\pm 1}\rho'(\sigma)=+\infty.\ee
It can be also checked that these somewhat exotic boundary conditions are consistent with the dynamics of the model (\ref{2}) since
they make sure that there is no flow of energy through the boundaries $\sigma=\pm 1$.

\section{Large $N$ limit of the Avan-Talon $r$-matrix}

Consider  an $r$-matrix  $r(N)$ given by the formula
\be r(N)_{12}=\sum_{k\neq l}^N\frac{\ri}{q_l-q_k}E_{kl}\otimes E_{lk}+\jp\sum_{k \neq l}^N\frac{\ri}{q_l-q_k}E_{kk}\otimes (E_{kl}-E_{lk}).\label{rn}\ee
Using the Poisson brackets (\ref{jef}), it is a matter of straightforward computation to verify   that the $r$-matrix (\ref{rn}) and the Lax matrix $L(N)$ satisfy  the following crucial property (cf.(\ref{stl}))
\be \{L(N)_1,L(N)_2\}_{P(N)}=-\frac{\ri N}{2} [r(N)_{12},L_1(N)]+\frac{\ri N}{2} [r(N)_{21},L_2(N)].\label{stln}\ee
We recall that it is this property (\ref{stln}) that  guarantees the $P(N)$-Poisson commutativity
of the traces $\tr L(N)^n$.  For a computational convenience, the  matrix $r(N)$  slightly  differs from the   $r$-matrix proposed by Avan and Talon in
\cite{AT}; in fact,  $r(N)$ is just another element   of a moduli space \cite{FP}  of all $r$-matrices verifying (\ref{stln}).  

\no Recalling the matrix $R(N)$ introduced in the previous section  it is easy to check that
\be [R(N)_1,r(N)_{12}]=-\ri\sum_{k\neq l}^N E_{kl}\otimes E_{lk};\label{321}\ee
\be [R(N)_2,r(N)_{12}]=\ri\sum_{k, l}^N E_{kl}\otimes E_{lk}  -\frac{\ri}{2} \biggl[\biggl(\sum_m^N E_{mm}\otimes E_{mm}\biggr),\biggr(\1_N\otimes \sum_{k,l}^NE_{kl} \biggr)\biggr]_+,\label{123}\ee
 where as usual $R(N)_1=R(N)\otimes 1$,  $R(N)_2=1\otimes R(N)$ and $[.,.]_+$ stands for an anticommutator. Now writing $R(N)$ as $Q_N(q(\si))$,
using the formulae (\ref{del}),(\ref{tif}) and (\ref{ddf}) and supposing that $r(N)$ can be written as a fuzzification  $Q_N(r)$ of some
function $r$ on $S^2\times S^2$,
the equations (\ref{321}) and  (\ref{123}) can be rewritten as 
\be [Q_N(q(\si))_1,Q_N(r)]=-\frac{2\ri}{N}Q_N\biggl(2\pi\delta(\si_1-\si_2)\delta(\phi_1-\phi_2) 
 -\delta(\si_1-\si_2)\biggr), \label{64}\ee
\be [Q_N(q(\si))_2,Q_N(r)]=\frac{2\ri}{N}Q_N\biggl(2\pi\delta(\si_1-\si_2)\delta(\phi_1-\phi_2)\biggr) -
\frac{\ri}{N}\biggl[ Q_N(\delta(\si_1-\si_2)),Q_N(2\pi\delta(\phi_2))\biggr]_+.\label{bcd}\ee
Using the fundamental quantization properties (\ref{p1}) and (\ref{p2}),  we deduce from (\ref{64}) and (\ref{bcd})  that $r$ must fulfil
\be \{q(\si_1), r (\si_1,\phi_1;\si_2,\phi_2)\}_B= -q'(\sigma_1) \partial_{\phi_1}r=\delta(\si_1-\si_2)(1-2\pi\delta(\phi_1-\phi_2)).\label{gog}\ee
\be \{q(\si_2), r (\si_1,\phi_1;\si_2,\phi_2)\}_B=-q'(\sigma_2) \partial_{\phi_2}r=  2\pi\delta(\si_1-\si_2)( \delta(\phi_1-\phi_2)- \delta(\phi_2)).\label{magog}\ee
 The relations (\ref{gog}) and (\ref{magog})  are indeed verified by the  following distribution 
 \be r(\si_1,\phi_1;\si_2,\phi_2)=-\frac{1}{ q'(\si_1)}\delta( \si_1-\si_2)\biggl(E(\phi_1-\phi_2)+E(\phi_2)\biggr).\label{rma}\ee
Here $E$ is viewed as a $2\pi$-periodic function on the  whole real axis $\bR$ which is given by the expression  $\phi-\pi{\rm sign}(\phi)$
when restricted to the interval  $[-\pi,\pi]$:
\be E(\phi):=\phi-\pi{\rm sign}(\phi), \quad  \phi\in [-\pi,\pi].\ee
The distribution (\ref{rma}) does coincide with the one found in \cite{Ho2}, moreover,
 the first order differential conditions  (\ref{gog}) and (\ref{magog})  determine   $r$ almost unambiguously.  The only ambiguity consists
in adding   a $\phi_1,\phi_2$-independent function  to  the solution (\ref{rma})  of  (\ref{gog}) and (\ref{magog}), however,  such   function  would be
a   large $N$ limit of  a bi-diagonal term which is absent in $r(N)$. We thus conclude
  that 
the classical Yang-Baxter  observable $r$ (\ref{rma})  of the totally classical Calogero model (\ref{2})
 is indeed the large $N$-limit  of the Avan-Talon $r$-matrix $r(N)_{12}$.

\section{Outlook}  As it is well-known, the Calogero model  has
attracted a lot of attention in pure  mathematics and in
mathematical physics (for reviews see \cite{Au,DKN,E,OP2,P,PO3,R})  since it  appears in a large variety of
contexts \cite{AB,BGS,BHV,BHKV,BGM,CLP,C,GT,GP,SC,H,I,LM,MP,PO2,SH,RMT,CMapp,Va1,Va2,W} ranging from condensed matter physics,  higher spin algebras, two-dimensional QCD or  fluid dynamics to microscopic description of black holes etc. We expect  a relevance of the totally classical
Calogero model in many of these contexts.  Speaking more generally, it would be interesting to construct  the totally
  classical models corresponding to various trigonometric, elliptic or even relativistic deformations of the Calogero one.   This
  would certainly contribute to a better understanding of  the   duality properties of the integrable systems. One can  also look for gauge transformations of 
  the $r$-functions in order to get rid of the $q$ dependence of $r$ (\ref{rma}), much  in the spirit of \cite{FP} which does that
  for the standard $r$-matrices.

\end{document}